\documentclass[11pt,a4paper]{article}
\pdfoutput=1

\usepackage[utf8]{inputenc}

\usepackage{jheppub}
\usepackage{bm}
\usepackage{bigints}

\usepackage{amsfonts}
\usepackage{amsmath}
\usepackage{bm,graphicx}
\usepackage{hyperref}
\usepackage{epsfig}
\usepackage{verbatim}   
\usepackage[subfigure]{graphfig}
\usepackage{epstopdf}
\usepackage{dcolumn}
\usepackage{color}
\usepackage{tabularx}
\usepackage{braket}
\usepackage{float}
\usepackage{tikz}
\usepackage{amssymb}
\usepackage[percent]{overpic}
\graphicspath{{./figures/}}
\usepackage{subfigure}
\usepackage{mwe}
\usepackage{soul}
\usepackage{ulem}
\usepackage{hhline}

\hypersetup{colorlinks, linkcolor = [rgb]{0,0.0,0.75}, citecolor = [rgb]{0,0.0,0.75}, urlcolor = [rgb]{0,0.0,0.75}}

\usepackage{soul}
\usepackage[utf8]{inputenc}
\usepackage{lineno}
\def\be{\begin{equation}}
\def\ee{\end{equation}}
\def\bea{\begin{eqnarray}}
\def\eea{\end{eqnarray}}

\newcommand{\nn}{\nonumber}
\newcommand{\la}{\label}

\definecolor{green}{rgb}{0,.5,0}

\newcommand{\remove}[1]{}

\begin{document}

\preprint{JLAB-THY-18-2792}

\author[a]{Raza Sabbir Sufian}
\author[b]{, Keh-Fei Liu}
\author[a]{, David G. Richards}
\affiliation[a]{Theory Center, Jefferson Lab, 12000 Jefferson Avenue, Newport News, VA 23606, USA}
\affiliation[b]{Department of Physics and Astronomy, University of Kentucky, Lexington, Kentucky 40506, USA}
\emailAdd{sufian@jlab.org }
\emailAdd{liu@pa.uky.edu}
\emailAdd{dgr@jlab.org }

\title{Weak Neutral Current Axial Form Factor Using $(\bar{\nu})\nu$-Nucleon Scattering and Lattice QCD Inputs}

\abstract{ We present a determination of the neutral current weak
  axial charge $G^Z_A(0)=-0.654(3)_{\rm stat}(5)_{\rm sys}$ using the
  strange quark axial charge $G^s_A(0)$ calculated with lattice QCD. We
  then perform a phenomenological analysis, where we combine the
  strange quark electromagnetic form factor from lattice QCD with
  (anti)neutrino-nucleon scattering differential cross section from
  MiniBooNE experiments in a momentum transfer region $0.24\lesssim
  Q^2 \lesssim 0.71$ GeV$^2$ to determine the neutral current weak
  axial form factor $G^Z_A(Q^2)$ in the range of $0\lesssim Q^2\leq 1$
  GeV$^2$. This yields a phenomenological value of
  $G^Z_A(0)=-0.687(89)_{\rm stat}(40)_{\rm sys}$. The value of
  $G^Z_A(0)$ constrained by the lattice QCD calculation of $G^s_A(0)$,
  when compared to its phenomenological determination, provides a
  significant improvement in precision and accuracy and can be used to
  provide a constraint on the fit to $G^Z_A(Q^2)$ for $Q^2>0$. This
  constrained fit leads to an unambiguous determination of
  (anti)neutrino-nucleon neutral current elastic scattering
  differential cross section near $Q^2=0$ and can play an important
  role in numerically isolating nuclear effects in this region.  We
  show a consistent description of $G^Z_A(Q^2)$ obtained from the
  (anti)neutrino-nucleon scattering cross section data requires a
  nonzero contribution of the strange quark electromagnetic form
  factor.  We demonstrate the robustness of our analysis by providing a post-diction of
  the BNL E734 experimental data.  }

\maketitle

\section{Introduction}
Precision measurements of various matrix elements associated with
(anti)neutrino-nucleon $(\bar{\nu})\nu-N$ scattering can directly
impact a wide variety of physical processes. These include
an understanding of solar
neutrino~\cite{Abe:2010hy,Bellini:2011rx,Gando:2010aa} and
atmospheric~\cite{Adamson:2011ig,Adamson:2011fa,Abe:2011ph} neutrino
oscillations, three non-vanishing mixing
angles~\cite{An:2012eh,Ahn:2012nd} resulting in a phase-violating CP
asymmetry leading to matter-antimatter asymmetry in the universe in
the three-neutrino framework, dynamics of neutron-rich core-collapse
supernovae~\cite{Bertulani:2009mf,Volpe:2013kxa}, axial-vector
structure of the nucleon, strange quark ($s$-quark) contribution
$G^s_A(0)\equiv \Delta s$ to the proton spin, and non-standard
interactions leading to beyond-the-standard-model
physics~\cite{Ohlsson:2012kf}. One such matrix element is the neutral
current weak axial charge $G^Z_A(0)$ and the corresponding
momentum-dependent form factor $G^Z_A(Q^2)$, arising through the
exchange of a neutral $Z^0$ boson between the lepton and
quarks.


In parity-violating $\vec{e}-p$ scattering experiments, the axial coupling to the nucleon, which is accessible through the parity-violating asymmetry, contains contributions beyond the tree-level $Z^0$ boson exchange. These contributions are important because in parity-violating scattering, the neutral weak axial form factor corresponding to this tree-level $Z^0$ boson exchange is suppressed by the weak charge of proton $(1-4\sin^2\theta_{\rm W})<<1$, where $\theta_{\rm W}$ is the Weinberg angle. In consequence, radiative corrections can have substantial contributions at higher orders of the strong coupling $\alpha_s$~\cite{Zhu:2000gn,Armstrong:2012bi}, and these contributions are poorly constrained by theory. Moreover, in parity-violating $\vec{e}-p$ scattering experiments at forward angles, the neutral current weak axial form factor is suppressed by an additional kinematic factor, while  the values of $G^Z_A(0)$ and $G^Z_A(Q^2)$ are not sufficiently constrained from these experiments at backward angles~\cite{Mueller:1997mt, Spayde:2003nr, Androic:2009aa, Baunack:2009gy,BalaguerRios:2016ftd}. For example, in the most recent determination~\cite{Androic:2018kni} of the weak charge $Q_{\text{weak}}$ of the proton, the effective weak axial charge  $G^{Z,{\rm eff}}_A(0)=-0.59(34)$ was determined using theoretical constraints from Ref.~\cite{Zhu:2000gn}. The analysis~\cite{Androic:2018kni} incorporated a dipole form~\cite{Hand:1963zz} for the isovector $G^{Z(T=1)}_A(Q^2)=[G^{Z,p}_A(Q^2)-G^{Z,n}_A(Q^2)]/2$ and isoscalar $G^{Z(T=0)}_A(Q^2)=[G^{Z,p}_A(Q^2)+G^{Z,n}_A(Q^2)]/2$ form factors where the superscripts $p$ and $n$ stand for proton and neutron, respectively. In contrast, the (anti)neutrino-nucleon $(\overline{\nu})\nu-N$ neutral current scattering process can be a perfect tool to extract $G^Z_A(Q^2)$ without these ambiguities of higher order radiative corrections in $\alpha_s$ and corrections associated with coherent strong interactions between intermediate particles, therefore serving as a complimentary tool to isolate the higher-order radiative corrections involved in the effective value of $G^{Z,\rm{eff}}_A(Q^2)$ in parity-violating $\vec{e}-p$ scattering experiments.

Unlike charged-current quasi-elastic (CCQE) scattering which is
sensitive only to the isovector current, neutral current elastic
$(\overline{\nu})\nu-N$ scattering is sensitive both to isoscalar and
isovector weak currents. In the absence of a precise knowledge of
$G^s_A(Q^2)$, the undetermined $G^Z_A(Q^2)$ is typically eliminated
from the $(\overline{\nu})\nu-N$ neutral current elastic scattering
data analysis by employing a value for $G^s_A(0)$ determined from
differing model assumptions or global analyses and a dipole
form~\cite{Hand:1963zz} with the dipole mass $M_{A,\rm{dip}}$ obtained
from the neutrino-nucleon CCQE scattering data analysis of isovector
axial form factor $G^{CC}_A(Q^2)$.

The major goal of this paper is to determine the neutral current weak
axial charge $G^Z_A(0)$ using first-principles lattice QCD, and then
obtain its $Q^2$-dependent form factor constrained using this charge.
We will demonstrate why a precise knowledge of $G^Z_A(Q^2)$ is vital
for the interpretation of neutrino scattering experiments, and show
that the strange quark contribution to nucleon electromagnetic form
factors cannot be ignored if we are to obtain a consistent value of
$G^Z_A(Q^2)$ from current $\overline{\nu}$ and $\nu$ neutral-current
scattering data. At the very lowest energies, far below that considered
here, the first example of
coherent elastic neutrino-nucleus scattering has recently been
observed~\cite{Akimov:2017ade}.
However, because of the nuclear effects and other possible systematics
associated with the published MiniBooNE
data~\cite{AguilarArevalo:2010cx,Aguilar-Arevalo:2013nkf} for
$(\overline{\nu})\nu-N$ neutral-current elastic-scattering
differential cross sections, and their use in the kinematic region
\mbox{$0.24\lesssim Q^2 \lesssim 0.71$ GeV$^2$} for our determination
of $G_A^Z(Q^2)$, we give post-dictions for the BNL E734
experiment~\cite{Horstkotte:1981ne,Ahrens:1986xe} differential cross
sections as a sanity check of the $G_A^Z(Q^2>0)$ extracted in this
analysis.  Finally, we show how the first-principles value for
$G^Z_A(0)$, and the $G^Z_A(Q^2)$ that we obtain, might provide a means to
expose and numerically quantify the Pauli blocking effect and the
binding energy of the carbon nucleus responsible for the fall-off of
the differential cross section in the low-energy region $Q^2\lesssim
0.15~{\rm GeV}^2$, and reinforces the need for a complete
first-principles calculation of the weak neutral-current axial vector
form factor.


\section{Determination of Neutral Current Weak Axial Form Factor}

An accurate determination of the $(\overline{\nu})\nu$ interaction with a free nucleon is vital to investigate nuclear effects in $(\overline{\nu})\nu$-nucleus scattering, and the effects of various nuclear model inputs in the  Monte-Carlo event generator of the scattering processes, and is needed to perform a comparison with the Standard Model physics. Along with the challenge of reconstructing the incoming neutrino beam energy $E_\nu$, these modern experiments~\cite{Horstkotte:1981ne,Ahrens:1986xe,Ahn:2001cq,Astier:2001yj,Lyubushkin:2008pe, Adamson:2007gu,Adamson:2008zt,Agafonova:2010dc, AguilarArevalo:2010zc,Aguilar-Arevalo:2013dva,AguilarArevalo:2010cx,Aguilar-Arevalo:2013nkf,Hiraide:2008eu, Anderson:2011ce,Fields:2013zhk,Fiorentini:2013ezn,Adams:2018fud,Acciarri:2015uup} face a defining challenge to systematically consider various nuclear effects in the initial- and final-state interactions, and  use a combination of various nuclear models  (for detailed discussion see the NuSTEC White Paper in Ref.~\cite{Alvarez-Ruso:2017oui}). While a lot of progress is being made by the nuclear physics community, as outlined in~\cite{Alvarez-Ruso:2017oui}, at this moment the community is not successful in verifying the proposed models in a quantitative sense (see also~\cite{Gallagher:2011zza,Formaggio:2013kya,Alvarez-Ruso:2014bla}). For example, the global analysis~\cite{Wilkinson:2016wmz} performed by T2K shows a comparison of results between different nuclear models implemented in the NEUT~\cite{Hayato:2009zz} neutrino interaction generator using the CCQE neutrino-nucleus scattering data of MiniBooNE~\cite{AguilarArevalo:2010zc, Aguilar-Arevalo:2013dva} and MINER$\nu$A~\cite{Fields:2013zhk,Fiorentini:2013ezn} experiments and observed significant differences between proposed models.


NUANCE~\cite{Casper:2002sd}, the Monte-Carlo simulation used by the MiniBooNE Collaboration, implemented neutral current elastic scattering off free nucleons based on Ref.~\cite{LlewellynSmith:1971uhs}, accounted for the production of intermediate pions~\cite{Rein:1982pf}, the dominance of Pauli-blocking at low $Q^2$, and included a relativistic Fermi gas model to account for bound states~\cite{Smith:1972xh}. Any outgoing pions in NUANCE simulations were given a 20\% probability to undergo final-state interaction. In the present analysis, we rely on the MiniBooNE published data of flux-integrated (anti)neutrino neutral current scattering differential cross sections in Refs.~\cite{AguilarArevalo:2010cx,Aguilar-Arevalo:2013nkf} to obtain $G^Z_A(Q^2)$. The MiniBooNE published flux-integrated differential cross sections are obtained through 
\bea
\frac{d\sigma_i^{\rm NCE}}{dQ^2} = \frac{\sigma_i/(\frac{dQ^2}{dT_N}\Delta T_N)}{N_N N_{\rm POT}\int \Phi_\nu dE_\nu}
\eea
where $\sigma_i$ is the number of entries for the the $i$-th bin of the unfolded nucleon kinetic energy distribution, $\frac{dQ^2}{dT_N}=2m_N$, $\Delta T_N=0.018$ GeV is the bin width of the unfolded nucleon kinetic energy distribution, $N_N=N_A\rho_{\rm oil} (4\pi R^3/3)$ is the number of nucleons in the detector with $N_A$ the Avogadro's number, $\rho_{\rm oil}$ is the density of the mineral oil and $R=610.6$ cm is the radius of the MiniBooNE detector. The number of protons on target is $N_{\rm POT} = 6.46\times10^{20}$, and $\int \Phi_\nu dE_\nu=5.22\times 10^{-10}$ cm$^{-2}\cdot$POT$^{-1}$ is the total integrated neutrino flux~\cite{MiniBooNE,Perevalov:2009zz}. In principle, instead of using $\int \Phi_\nu dE_\nu$, one can calculate the differential cross section bin-by-bin using the MiniBooNE flux in the data release~\cite{MiniBooNE} and then obtain the differential cross section. One can check that one obtains comparable results within uncertainties whether one uses the flux-integrated cross sections, or those obtained bin-by-bin using the posted flux tables. We also note that the reported differential cross section is given in terms of quasi-elastic momentum transfer $Q^2_{\rm QE}$ and the interpretation of $Q^2_{\rm QE}$ in terms of the assumed momentum transfer from the neutrino introduces additional model dependence which is beyond the scope of this manuscript. Since the main goal of this paper is to show the effect of the lattice QCD input of strange-quark electromagnetic form factors and the strange quark axial charge in understanding and controlling a few of the many systematic uncertainties in the data analysis of the neutral current neutrino scattering experiments rather than a detailed discussion of uncertainty propagation in neutrino experimental data, we use the MiniBooNE published data of the flux-integrated differential cross sections in the following analysis.


In this section, we first present a direct determination of the neutral current weak axial charge using the  relation between the charged current axial form factor $G^{CC}_A(Q^2)$ and strange quark axial form factor~\cite{Garvey:1993sg,Garvey:1992cg}
\bea \la{gzaeq}
G^Z_A(Q^2) &=& \frac{1}{2} [-G^{CC}_A(Q^2) + G^s_A(Q^2)] .
\eea
At $Q^2=0$, we can directly calculate the neutral current weak axial charge $G^Z_A(0)$ using the experimental value of $G^{CC}_A(0)=g_A=1.2723(23)$~\cite{PDG} and the lattice QCD calculation of the strange quark axial charge obtained in Ref.~\cite{Liang:2018pis} which is shown to satisfy the anomalous Ward identity. Three lattice ensembles were used, and the value  
\bea \la{gs0lat}
G^s_A(0)\equiv \Delta s=-0.035(6)_{\rm stat}(7)_{\rm sys}
\eea
 was obtained in the continuum limit and physical pion point through a simultaneous fit in lattice spacing, volume and pion mass~\cite{Liang:2018pis}. This value is consistent with but more precise than the first simultaneous extraction of spin-dependent parton distributions and fragmentation functions from a global QCD analysis in Ref.~\cite{Ethier:2017zbq} where the authors obtained $G^s_A(0)=-0.03(10)$. We can now use the lattice QCD estimate of  $G^s_A(0)$ to obtain
\bea \la{gz0lat}
G^Z_A(0) &=& \frac{1}{2} [-G^{CC}_A(0) + G^s_A(0)]\nn \\
&=& -0.654(3)_{\rm stat}(5)_{\rm sys} .
\eea
$G^Z_A(0)$ obtained this way is also independent of the systematics involved in the extrapolation of the $G^Z_A(Q^2)$ data using the $z$-expansion fit presented below. This value of $G^Z_A(0)$ is quite precise and is free from systematics associated with nuclear effects and the other systematics that are involved in the neutrino scattering on a nuclear target. This is a good example of how lattice QCD calculations can benefit our understanding of low-energy nuclear physics. For example, for the first time, the Flavour Lattice Averaging Group (FLAG) included several nucleon quantities calculated using lattice QCD in their 2019 review~\cite{Aoki:2019cca}, including the calculation of the strange quark contribution to nucleon electromagnetic form factors~\cite{Sufian:2016pex} to be used in the following calculation. It is also important to note that, with significant developments in numerical techniques, lattice QCD calculations can now calculate $G^{CC}_A(0)$  with  uncertainties  at  the  few percent level~\cite{Chang:2018uxx,Gupta:2018qil,Alexandrou:2019brg} and future improved lattice QCD calculations of  $G^{CC}_A(Q^2)$ and nucleon electromagnetic form factors with fully controlled systematic uncertainties will have direct impact on the understanding  of the neutrino-nucleus scattering~\cite{Kronfeld:2019nfb}.

In the absence of first-principles calculations of the $Q^2$-dependent
charged-current and strange quark axial form factors with all
systematic errors under control, in this phenomenological analysis, we
use MiniBooNE differential cross section data of the
$(\overline{\nu})\nu-N$ neutral current elastic
scattering~\cite{AguilarArevalo:2010zc, Aguilar-Arevalo:2013dva} to
extract $G^Z_A(Q^2)$. The $(\overline{\nu})\nu-N$ neutral current
elastic differential cross-section $d\sigma/dQ^2$, assuming
conservation of the vector current~\cite{Feynman:1958ty} which equates
the vector form factors in the electromagnetic interaction to the
corresponding form factors in the weak interaction, with $Q^2$
dependence implicit in the form factors, can be written
as~\cite{Garvey:1993sg,Garvey:1992cg}: \bea \la{dsigdq}
\frac{d\sigma_{\nu(\overline{\nu})N\to \nu(\overline{\nu})N}}{dQ^2} =
\frac{G_F^2}{2\pi}\frac{Q^2}{E_\nu^2} (A\pm BW+CW^2), \eea where \bea
\la{defx} A\!&=&\!\frac{1}{4}\![(G^Z_A)^2\!(1\!+\!\tau)\! -
  \!\{\!(F_1^Z)^2 \!-\! \tau(F_2^Z)^2\}\!(1\! -\! \tau)\!+ \!4\tau
  F_1^Z \!F_2^Z],\nn\\ B&=&-\frac{1}{4}G^Z_A(F_1^Z+F_2^Z), \nn\\ C&=&
\frac{1}{64\tau} [(G^Z_A)^2+(F^Z_1)^2+\tau
  (F_2^Z)^2],\nn\\ W&=&4(E_\nu/M_p-\tau), \eea and the $+(-)$ sign is
for $\nu(\overline{\nu})$ scattering off a free nucleon. Here $G_F$ is
the Fermi constant~\cite{PDG}, $E_\nu$ is the neutrino average beam
energy, $M_p$ is the nucleon mass, and $\tau=Q^2/4M_p^2$.  The weak
neutral current Dirac and Pauli form factors $F_{1,2}^Z(Q^2)$ in
Eq.~(\ref{defx}) can be calculated in terms of the proton and neutron
electromagnetic form factors $F_{1,2}^{p,n}(Q^2)$ and strange quark
form factors $F_{1,2}^s(Q^2)$ \bea\la{WNCFF} F_{1,2}^Z(Q^2)&&=
\big(\frac{1}{2}-\sin^2\theta_{\rm W}\big)
\big(F_{1,2}^p(Q^2)-F_{1,2}^n(Q^2)\big)\nn \\ &&-\sin^2\theta_{\rm
  W}\big(F_{1,2}^p(Q^2)+F_{1,2}^n(Q^2)\big) -
\frac{F_{1,2}^s(Q^2)}{2}. 
\eea 
To calculate $F_{1,2}^Z(Q^2)$, we
use the most precise values of $F^s_{1,2}(Q^2)$ obtained from the
lattice QCD
calculations~\cite{Sufian:2016pex,Sufian:2016vso,Sufian:2017osl} at
the physical pion mass and in the continuum and infinite-volume
limits; we note that some other lattice groups, using different
  discretizations of the fermion action, but without such
  extrapolations, obtain smaller values for the characteristic strange-quark magnetic
  moments~\cite{Green:2015wqa,Djukanovic:2019jtp,Alexandrou:2019olr} than the value
  $G^s_M(0) = -0.064(14)(09) \mu_N$~\cite{Sufian:2016pex} that
  we use
  here. For $F^{p,n}_{1,2}(Q^2)$, we use the most
  recent model-independent $z$-expansion
  fit~\cite{Hill:2010yb,Epstein:2014zua}, including
  two-photon-exchange corrections, to world electron-scattering
  experimental data from Ref.~\cite{Ye:2017gyb}. With $F_{1,2}^Z(Q^2)$
  determined this way, we use $d\sigma/dQ^2$ from the MiniBooNE
  experiments~\cite{AguilarArevalo:2010cx,Aguilar-Arevalo:2013nkf}
  over a range of $Q^2$ discussed below to extract $G^Z_A(Q^2)$ from
  Eq.~(\ref{dsigdq}).  It is worth mentioning that, a somewhat similar
  approach was taken in Ref.~\cite{Pate:2003rk} to obtain strange
  quark Sachs electromagnetic form factors $G^s_{E,M}(Q^2)$ and
  $G^s_A(0)$.


MiniBooNE used a mineral-oil based (CH$_2$)  Cherenkov detector, thereby permitting $(\overline{\nu})\nu$ scattering from both bound protons and neutrons in carbon (C), and from free protons in hydrogen (H).  To obtain $(\overline{\nu})\nu-N$-scattering off free nucleons, different efficiency corrections $\eta$ associated with neutral current elastic scattering on free protons $(p)$ in H and on bound protons(neutrons) $p(n)$ in carbon are combined as:
\bea\la{eff}
\frac{d\sigma_{\nu(\overline{\nu})N\to \nu(\overline{\nu})N}}{dQ^2} &=& \frac{1}{7} \eta_{\nu(\overline{\nu})p,H} (Q^2)\frac{d\sigma_{\nu(\overline{\nu})p\to \nu(\overline{\nu})p,H}}{dQ^2}\nn \\
&+&\frac{3}{7} \eta_{\nu(\overline{\nu})p,C} (Q^2)\frac{d\sigma_{\nu(\overline{\nu})p\to \nu(\overline{\nu})p,C}}{dQ^2}\nn \\
&\!+\!&\frac{3}{7} \eta_{\nu(\overline{\nu})n,C} (Q^2)\!\frac{d\sigma_{\nu(\overline{\nu})n\to \nu(\overline{\nu})n,C}}{dQ^2}.
\eea
The efficiency corrections $\eta_{\nu(\overline{\nu}) p,H}$, $\eta_{\nu(\overline{\nu}) p,C} $, $\eta_{\nu(\overline{\nu}) n,C} $  for each type of neutral current scattering process are estimated from Monte Carlo simulation as a function of $Q^2$. These are defined as the ratios of the efficiency for a particular type of neutral current elastic event to the average efficiency of all the neutral current elastic events in bins of $Q^2$. To minimize effects of possible unknown systematics related to the $\eta$-values, we restrict ourselves to data in
the $Q^2$-regions where all three $\eta$'s for each $\nu$ or $\overline{\nu}$ in Eq.~\eqref{eff} are equal to 1 within about 2\%.  Moreover, the statistical uncertainty of the lattice QCD calculation of $F_{1,2}^s(Q^2)$ limits our ability to reliably go beyond $Q^2=0.75$ GeV$^2$. Therefore, for the determination of $G^Z_A(Q^2)$, we consider $d\sigma/dQ^2$ data extracted by MiniBooNE~\cite{AguilarArevalo:2010cx,Aguilar-Arevalo:2013nkf} only in the regions $0.24 \lesssim Q^2\lesssim 0.71$ GeV$^2$  (for $\overline{\nu}-N$ scattering) and $0.37 \lesssim Q^2\lesssim 0.71$ GeV$^2$ (for $\nu-N$ scattering).

Since the incoming neutrino beam energy $E_\nu$ cannot be directly measured, to estimate the $Q^2$ for the neutral current elastic scattering, MiniBooNE reconstructed $Q^2$ by calculating the total kinetic energy of the final state nucleons and expressed the reconstructed $Q^2$ within bin boundaries with a bin size of about $0.068$ GeV$^2$.  To take into account the uncertainty of the reconstructed $Q^2$, we assume a normal distribution with the central value at the middle of the bin and standard deviation half of the bin size.  The error bars presented in the MiniBooNE differential cross section data include the total normalization error due to both systematic and statistical errors, and these are used in our analysis. The systematic errors are correlated and common to both $\overline{\nu}$ and $\nu$-scattering measurements by MiniBooNE and included in the uncertainties of the differential cross sections as mentioned in~\cite{AguilarArevalo:2010cx,Aguilar-Arevalo:2013nkf}.

With $G^Z_A(Q^2)$ obtained from the combination of experimental and lattice QCD data in the \mbox{$0.24 \lesssim Q^2\lesssim 0.71$ GeV$^2$} region as described above, we perform a $z$-expansion fit~\cite{Hill:2010yb,Epstein:2014zua}:
\bea \la{zexp}
G^{Z,z-\text{exp}}_{A}(Q^2) = \sum^{k_\text{max}}_{k=0} a_k z^k, \nn \\
z=\frac{\sqrt{t_{\text{cut}}+Q^2}-\sqrt{t_{\text{cut}}}}{\sqrt{t_{\text{cut}}+Q^2}+\sqrt{t_{\text{cut}}}}
\eea 
 to the $G^Z_A(Q^2)$ data to obtain the neutral current weak axial charge $G^Z_A(0)$. In our fit to $G_Z^A(Q^2)$, we take into account the correlations between the lattice QCD data for $F_{1,2}^s(Q^2)$ at different $Q^2$, but do not take account of possible correlations between the MiniBooNE data at different $Q^2$ bins.  If instead we perform a fully uncorrelated fit, the resulting value of $G^Z_A(0)$ is within 5\% of the value obtained in the above fit, and as expected the uncertainty is smaller, by a factor of around three. We use $t_{\text{cut}} = (3m_\pi)^2$, representing the leading three-pion
threshold for states that can be produced by the axial current.  As we increase the number of fit parameters, the uncertainties in the higher order coefficients in $z$-expansion increase. However, $a_0=G^Z_A(0)$ remains the same within the uncertainty irrespective of the higher order terms in the fit. This means that the higher order terms ($k\geq 2$) do not have significant impact on the fit. We consider the $z$-expansion fit with 4 terms ({\it{i.e.}} $k_{\rm max}=3$) for the subsequent analysis and add the differences in the central values between the 2, 3, and 4-term fits in quadrature as the systematic uncertainty of the fit to obtain a final value
\bea \la{gz0}
G^Z_A(0) = -0.687(89)_{\rm stat}(40)_{\rm sys}
\eea
where the uncertainties in the parentheses are from statistics and systematics respectively. We see that this value of $G^Z_A(0)$ in Eq.~\eqref{gz0} is statistically consistent with that quoted in Eq.~\eqref{gz0lat} using the lattice QCD calculation of $G^s_A(0)$ in Ref.~\cite{Liang:2018pis}. However, the uncertainty of $G^Z_A(0)$ in Eq.~\eqref{gz0} is almost sixteen times larger compared to the uncertainty of  $G^Z_A(0)$ obtained in Eq.~\eqref{gz0lat}. We list the fit parameters in the first block of Table~\ref{table:r0}.
\begin{table*}[t]
  \centering
  \begin{tabular}{|c|c|c|c|c|}
  \hline
    $k_{\rm max}$  & Fit parameters  &$a_0=G^Z_A(0)$ &  $\chi^2/{\rm d.o.f.}$\\
    \hline
    1 & $a_1=1.29(10)$& -0.726(30) & 0.63 \\
    \hline
    2 & $a_1=0.88(43), a_2=0.79(72)$& -0.678(65) & 0.20\\
    \hline
    3 & $a_1=0.97(52), a_2=0.41(88), a_3=0.54(1.54)$ &  -0.687(89) & 0.15\\
     \hhline{|=|=|=|=|}
     \vtop{\hbox{\strut $k_{\rm max}$ [$G^s_A(0)$ from}  \hbox{\strut  lattice QCD]}} & Fit parameters  &$a_0=G^Z_A(0)$ & $\chi^2/{\rm d.o.f.}$ \\
    \hline
    3 & $a_1=0.74(18), a_2=0.80(1.06), a_3=0.49(1.67)$ & -0.654(06) & 0.16 \\
    \hline
  \end{tabular}
  \caption{Parameters of $z$-expansion fit to Eq.(~\ref{zexp}) for $G^Z_A(Q^2)$ with $2$, $3$, and $4$ terms.  The $z$-expansion fit parameters with and without the lattice QCD input of strange quark contribution to nucleon spin is shown for comparison.  }
  \label{table:r0}
\end{table*}

From Eq.~\eqref{gzaeq} the strange quark axial form factor can be written as
\bea \la{gsA}
G^s_A(Q^2)=2G^Z_A(Q^2)+G^{CC}_A(Q^2).
\eea
With {\mbox{$G^{CC}_A(0)=g_A=1.2723(23)$~\cite{PDG}}} and $G^Z_A(0)$ from Eq.~(\ref{gz0}), we obtain
\bea\la{deltas}
G^s_A(0) \equiv \Delta s= -0.102 (178)_{\rm stat}(80)_{\rm sys}.
\eea
The comparison between the uncertainties in $G^s_A(0)$ values obtained in Eq.~\eqref{deltas} and in the lattice calculation of Eq.~\eqref{gs0lat} in~ Ref.~\cite{Liang:2018pis} demonstrates the impact of a precise lattice QCD calculation of $G^s_A(0)$ for a  better understanding of the neutral current scattering processes.

The large statistical uncertainty  in $G^s_A(0)$ in Eq.~(\ref{deltas}) is understood qualitatively through error propagation arguments arising from the cancellation of two large numbers. That said, one important feature of this analysis is that the $(\overline{\nu})\nu-N$  neutral current elastic cross section depends directly on the strange quark contribution, and therefore no assumptions about SU(3) flavor symmetry or fragmentation functions  is needed to obtain $G^s_A(0)$. We direct the reader to the discussion of the influences of SU(3) flavor symmetry in Ref.~\cite{Ethier:2017zbq} and fragmentation functions in Ref.~\cite{Leader:2010rb,Leader:2011tm}. Within the uncertainty, $G^s_A(0)$ in Eq.~(\ref{deltas}) is consistent with  $G^s_A(0)\sim -0.1$ obtained in Refs.~\cite{  Alexakhin:2006oza,Airapetian:2006vy,deFlorian:2009vb,Hirai:2008aj,Blumlein:2010rn,Nocera:2014gqa,Leader:2014uua,Sato:2016tuz}, {\mbox{$G^s_A(0)=0.08(26)$}} from MiniBooNE $\nu-N$  neutral current elastic scattering~\cite{AguilarArevalo:2010cx}, and {\mbox{$G^s_A(0)=0, -0.15(7), -0.13(09),-0.21(10)$}} [for various values of $G^s_{M}(0)$] from BNL E734 analysis~\cite{Garvey:1992cg}.  

We also fit the $G^Z_A(Q^2)$ data  using the dipole form~\cite{Hand:1963zz}
\bea
G^Z_A(Q^2)= \frac{G^Z_A(0)}{1+\bigg(\frac{Q^2}{M^2_{A,\rm{dip}}}\bigg)^2}
\eea 
with the $G^Z_A(0)$ value in Eq.~\eqref{gz0lat} and obtain
\bea
M_{A,\rm{dip}} = 1.057(14)~{\rm GeV}
\eea
with $\chi^2/{\rm d.o.f.}=0.71$. This value of $M_{A,\rm{dip}}=1.057(14) \rm{GeV}$ is consistent with the world average $M_{A,\rm{dip}}\sim 1$ GeV~\cite{Bernard:2001rs}. We emphasize that we do not use $G^Z_A (Q^2 )$ obtained from the dipole fit in our subsequent analysis and present this result only for the purpose of qualitative illustration.

\begin{figure}[htbp]
  \centering
  \subfigure[]{\includegraphics[height=6.5cm,width=9.5cm]{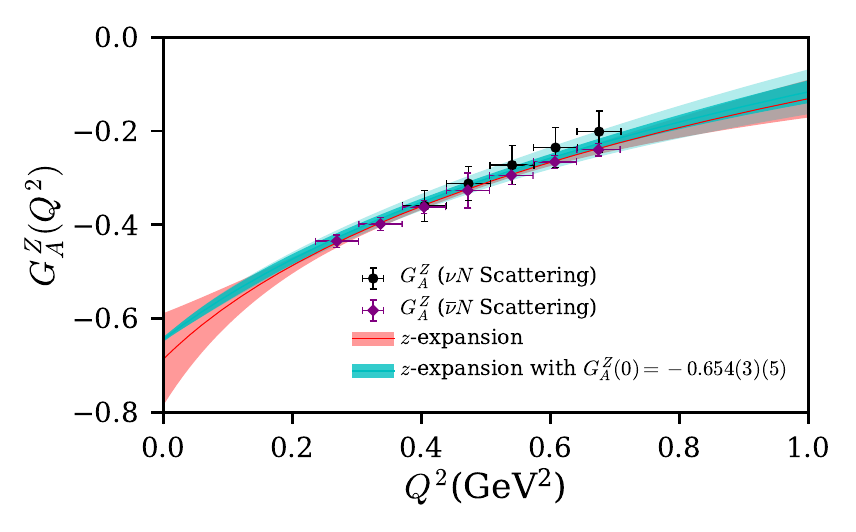}\la{fig1a}}
  \subfigure[]{\includegraphics[height=6.3cm,width=9.3cm]{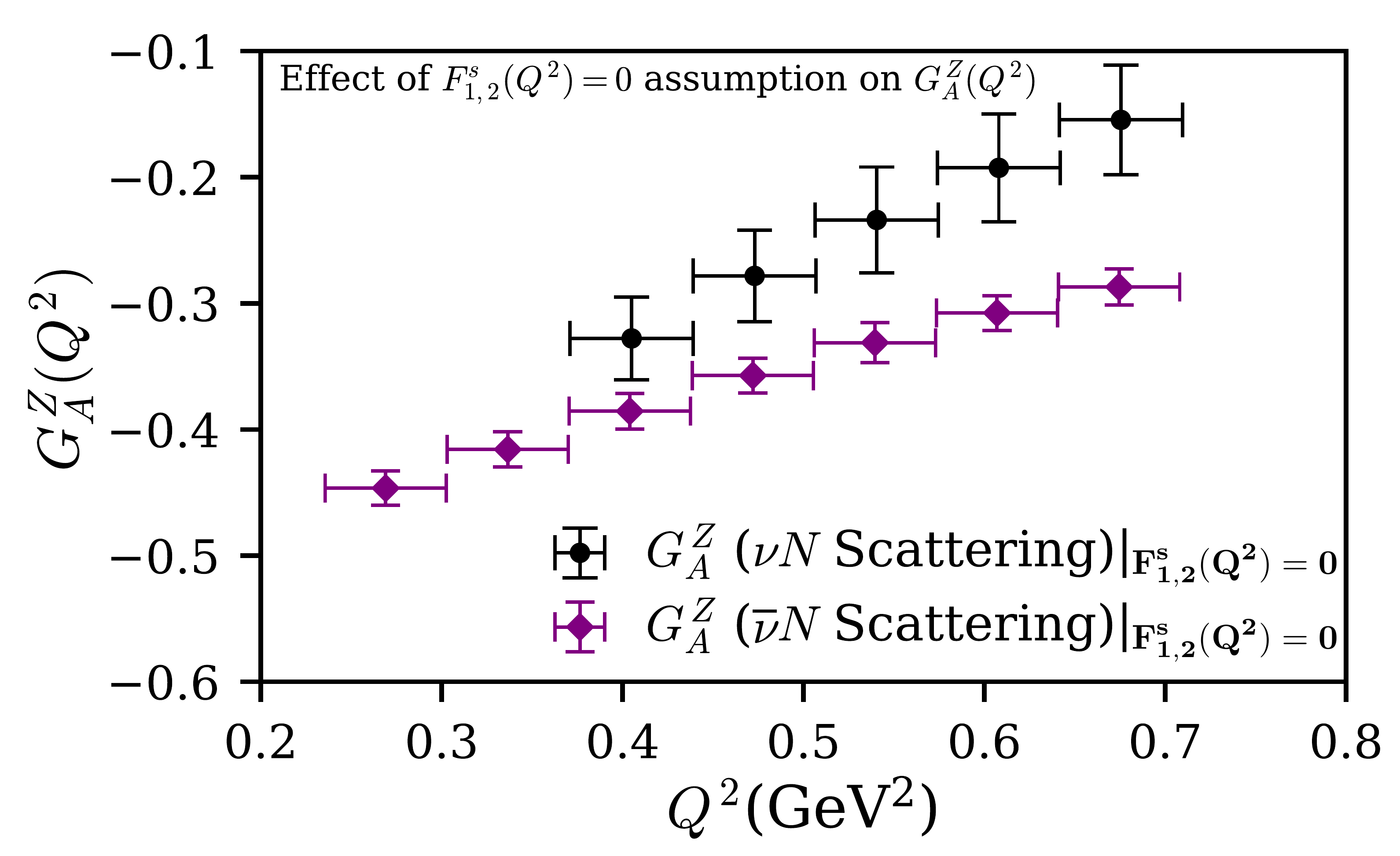}\la{fig1b}}
  \caption{\label{fig1} Fig.~\ref{fig1a}: Neutral current weak axial form factor $G^Z_A(Q^2)$ obtained from analyses combining MiniBooNE data of $(\overline{\nu})\nu-N$ scattering differential cross sections, lattice QCD estimates of strange quark  electromagnetic form factors, strange quark axial charge and a model-independent fit of the nucleon electromagnetic form factors  experimental data used from Ref.~\cite{Ye:2017gyb}. The red band shows 4-term $z$-expansion fit to the $G^Z_A(Q^2)$ data where the systematics coming from 2-terms, 3-terms $z$-expansion fit are added in quadrature to the  statistical uncertainty of the 4-term fit. The cyan band shows the $z$-expansion fit to the $G^Z_A(Q^2)$ data  when $G^s_A(0)$ is fixed by the value in Eq.~\eqref{gs0lat}.  The outer cyan band indicates an estimate of additional systematic uncertainty from the strange quark electromagnetic form factors calculated in other lattice QCD calculations~\cite{Green:2015wqa,Djukanovic:2019jtp,Alexandrou:2019olr} discussed in the text.       Fig.~\ref{fig1b}: Neutral current weak axial form factor $G^Z_A(Q^2)$  by assuming zero strange quark electromagnetic form factor, {\it i.e.} $F^s_{1,2}(Q^2)=0$. An assumption of  $F^s_{1,2}(Q^2)=0$ leads to different values of neutral current weak axial form factor of the nucleon at the same $Q^2$, obtained from $\overline{\nu}-N$ and $\nu-N$ neutral current scattering differential cross sections data in Refs.~\cite{AguilarArevalo:2010cx,Aguilar-Arevalo:2013nkf}.}
\end{figure}

We present the results of the above analysis of $G^Z_A(0)$ and $G^Z_A(Q^2>0)$ from the $z$-expansion fits  in Fig.~\ref{fig1a} and list the corresponding fit parameters in Table~\ref{table:r0}. The second block in Table~\ref{table:r0} presents the fit to $G^Z_A(Q^2)$ where we constrain $G^Z_A(0)$ by the value of $G^s_A(0)$ presented in Eq.~\eqref{gz0lat}. The $z$-expansion fit with and without this constraint are illustrated in Fig.~\ref{fig1a} and the low-$Q^2$ uncertainty is reduced significantly with this constraint.

An important result, demonstrated in Fig.~\ref{fig1b}, is that, although the $F^s_{1,2}(Q^2)$ contribution to the nucleon is much smaller than the valence quark contribution as shown in Refs.~\cite{Sufian:2016pex,Sufian:2016vso,Sufian:2017osl}, the assumption of \mbox{$F^s_{1,2}(Q^2)=0$} in Eqs.~\eqref{dsigdq}-\eqref{WNCFF} will lead to different results for the nucleon matrix element $G^Z_A(Q^2)$ at the same value of $Q^2$ obtained from the $\overline{\nu}$ and the $\nu$ scattering cross-section data. That is, if the contributions of $F_{1,2}^s(Q^2)$ in Eq.~\eqref{WNCFF} are ignored, one obtains $G^Z_A(Q^2)$-values from the MiniBooNE neutrino and antineutrino scattering differential cross section data that are not the same within their uncertainties. For example, using the $F^s_{1,2}(Q^2)$ values from~\cite{Sufian:2016pex,Sufian:2016vso,Sufian:2017osl}, we obtain $G^Z_A=-0.360(33)$ and $G^Z_A=-0.362(14)$ at $Q^2=0.405\, {\rm GeV}^2$; and $G^Z_A=-0.202(44)$ and $G^Z_A=-0.240(13)$ at $Q^2=0.675\, {\rm GeV}^2$, from the $\nu-N $ and $\overline{\nu}-N$ scattering data, respectively. On the other hand, with an assumption of $F^s_{1,2}(Q^2)=0$, we obtain $G^Z_A=-0.328(33)$ and $G^Z_A=-0.385(14)$ at $Q^2=0.405\, {\rm GeV}^2$; and $G^Z_A=-0.154(43)$ and $G^Z_A=-0.287(14)$ at $Q^2=0.675\, {\rm GeV}^2$ from the $\nu-N $ and $\overline{\nu}-N$ scattering data, respectively. Comparing the $G^Z_A(Q^2)$ values shown in Fig.~\ref{fig1a} and Fig.~\ref{fig1b} we find that $G^Z_A(Q^2)$ extracted from the MiniBooNE $\nu-N $scattering data~\cite{AguilarArevalo:2010cx} shift towards smaller negative values and those extracted from the MiniBooNE $\overline{\nu}-N$ scattering data~\cite{Aguilar-Arevalo:2013nkf} shift towards larger negative values when $F^s_{1,2}$ form factors are set to zero. This discrepancy can be clearly seen in Fig.~\ref{fig1b}. This exercise shows that the contribution of $F^s_{1,2}$ cannot be ignored, as has mostly been the case in previous such calculations, and that a precise lattice QCD input of $F^s_{1,2}(Q^2)$ plays a critical role in the understanding of neutral current (anti)neutrino-nucleon scattering.

 Comparing with other lattice QCD calculations of the strange quark electromagnetic form factors~\cite{Green:2015wqa,Djukanovic:2019jtp,Alexandrou:2019olr}, we note that the $G^s_E(Q^2)$ obtained in these calculations are in statistical agreement with that determined in~\cite{Sufian:2016vso,Sufian:2017osl}. However, the $G^s_M(Q^2)$ obtained in these calculations~\cite{Green:2015wqa,Djukanovic:2019jtp,Alexandrou:2019olr} are almost two times smaller in their central values compared to $G^s_M(Q^2)$ determined in~\cite{Sufian:2016vso,Sufian:2017osl} in the $0.24\lesssim Q^2 \lesssim 0.71$ GeV$^2$-region where the lattice QCD values of $G^s_{E,M}(Q^2)$ have been used, as shown in Fig.~\ref{fig1a}. It is seen from Fig.~\ref{fig1b} that a relatively larger magnitude of $G^s_{E,M}(Q^2)$ is favored to obtain statistically consistent $G^Z_A(Q^2)$ from the $\nu-N $ and $\overline{\nu}-N$ scattering data. We estimate the systematic uncertainty in the calculation of $G^Z_A(Q^2)$ using $G^s_M(Q^2)$ in the $0.24\lesssim Q^2 \lesssim 0.71$ GeV$^2$-region obtained in other lattice QCD calculations~\cite{Green:2015wqa,Djukanovic:2019jtp,Alexandrou:2019olr}, and indicate it as the outer cyan band shown in Fig.~\ref{fig1a}.

\section{Reconstruction of MiniBooNE Differential Cross Sections}

\begin{figure}[htp]
  \centering
  \subfigure[]{\includegraphics[height=6.8cm,width=9.5cm]{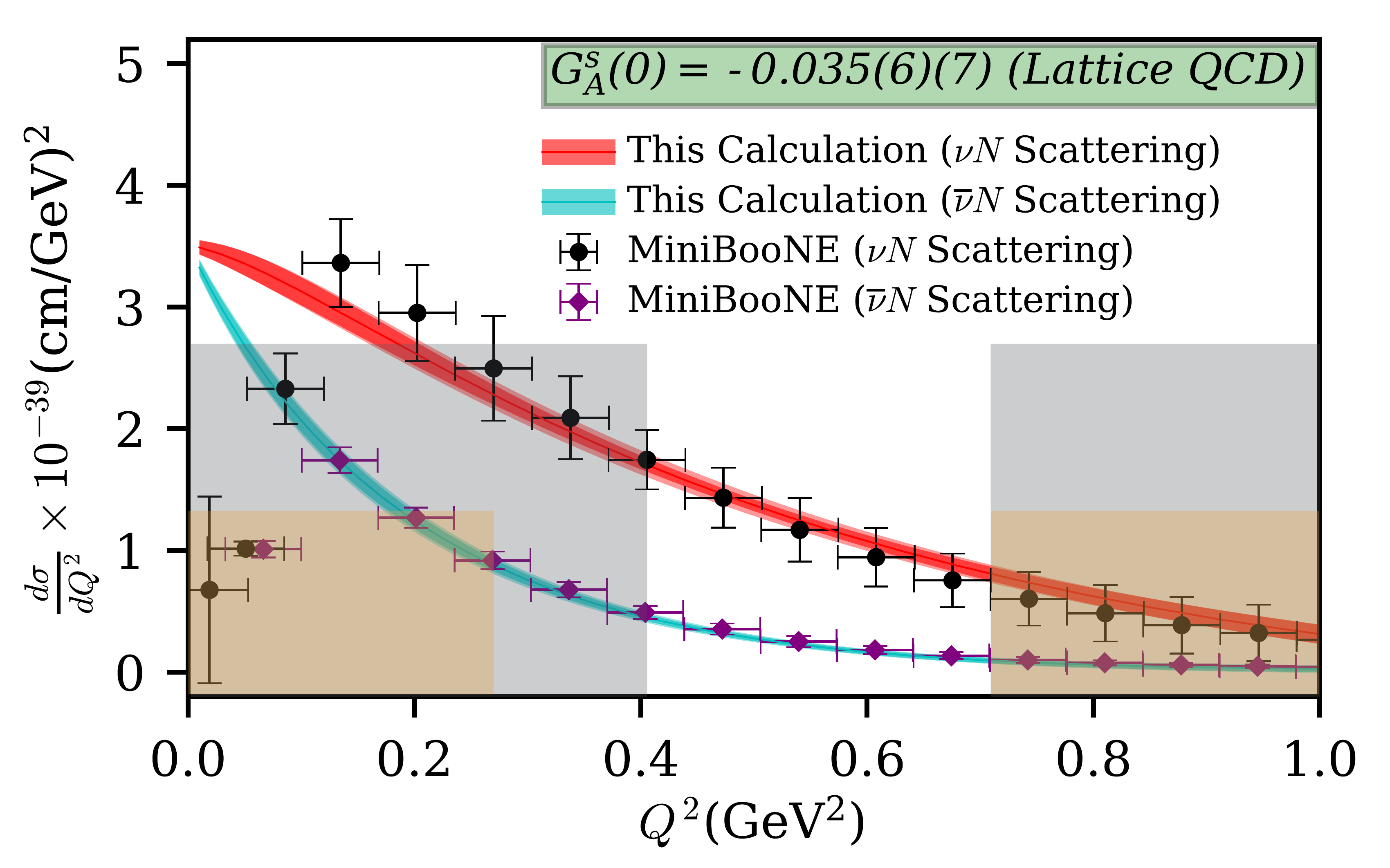}\label{3a}}
  \subfigure[]{\includegraphics[height=6.8cm,width=9.5cm]{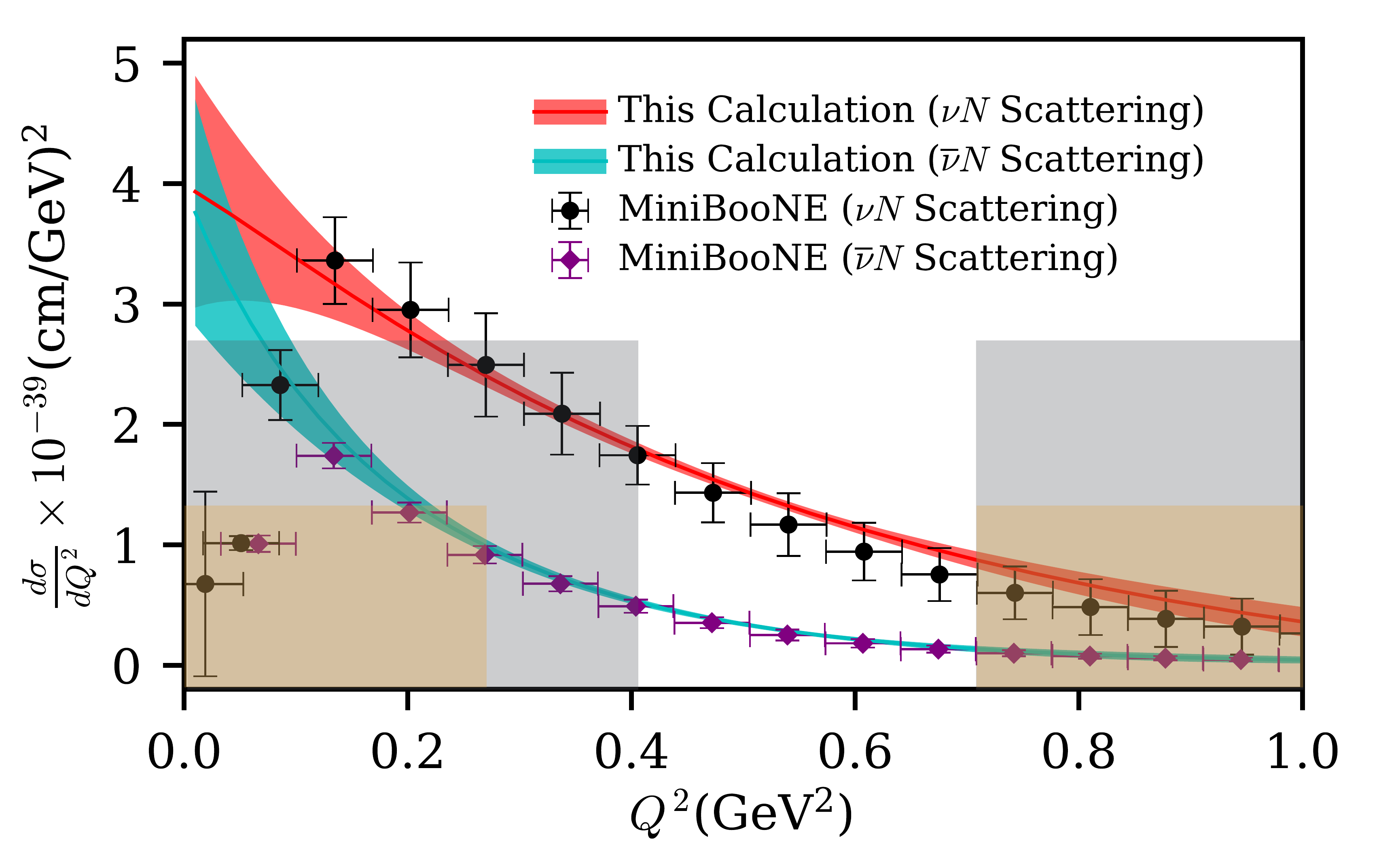}\label{3b}}
  \caption{\label{matelem}
   Comparison of the  $(\overline{\nu})\nu-N$ differential cross section between our  analysis and MiniBooNE extractions. Fig.~\ref{3a} shows the differential cross sections obtained using the fit parameters of $G^Z_A(Q^2)$ from Table~\ref{table:r0}  when lattice QCD estimate of $G^s_A(0)$ is used to constrain $G^Z_A(0)$ in the $z$-expansion fit. The uncertainty coming from neutrino beam energy $\delta E_\nu=\pm 0.1$ GeV is added as a systematic error in the $d\sigma / dQ^2$ extraction and shown in the outer light red and light cyan bands in Fig.~\ref{3a}. Fig.~\ref{3b} shows the differential cross sections obtained using the $z$-expansion fit results presented in Table~\ref{table:r0} without using the direct determination $G^Z_A(0)$ using lattice $G^s_A(0)$ in the fit.  The shaded gray and orange indicate the value of $Q^2$ excluded in our determination of $G^Z_A(Q^2)$ for the  $\nu-N$ and $\overline{\nu}-N$ scattering data, respectively. The lowest four $Q^2$ data points for $\nu-N$ scattering are compiled from Ref.~\cite{Perevalov:2009zz}. }
   \la{fig3}
\end{figure}

With our knowledge of $G^Z_A(Q^2)$, Eq.~(\ref{defx}) can now be used to obtain the $(\overline{\nu})\nu-N$ differential cross sections in the full $0\lesssim Q^2\leq 1$ GeV$^2$ kinematic region as shown in Fig.~\ref{fig3}. We are able to successfully reconstruct the MiniBooNE data outside the $Q^2$-region that was used for the determination of $G^Z_A(Q^2)$. It is evident from Fig.~\ref{fig3} that in $Q^2\lesssim 0.15$ GeV$^2$, the free-nucleon scattering prediction starts to deviate from the MiniBooNE result. One reason is the Pauli blocking effect for which low-momentum transfer interactions are suppressed due to occupied phase space. This effect was already included in the NUANCE Monte-Carlo simulation and shown to have an impact exactly in the $Q^2\lesssim 0.15$ GeV$^2$ region~\cite{Katori,AguilarArevalo:2007ab}. A further possible reason is nuclear shadowing which is related to the phenomenon that, at low $Q^2$, the resolution is not sufficient to resolve a single nucleon wave function and therefore the differential cross section $d\sigma / dQ^2$ decreases~\cite{Katori:2014qta}. We use the average beam energy  $E_\nu=0.80$ GeV and \mbox{$E_{\overline{\nu}}=0.65$ GeV} in Eq.~\eqref{dsigdq} for the neutrino and anti neutrino scattering experiments respectively as mentioned in the MiniBooNE papers~\cite{AguilarArevalo:2010cx,Aguilar-Arevalo:2013nkf}. To investigate the effect of $E_\nu$ in the calculation of $(\overline{\nu})\nu-N$ differential cross sections in Eq.~\eqref{dsigdq}, we assign a $\delta E_\nu=\pm 0.1$ GeV uncertainty in the values of $E_\nu$ and examine its effect on the extracted $d\sigma / dQ^2$ in our analysis. We see a variation of $E_\nu$ by more than an amount of 10\% does not lead to a significant difference in the $d\sigma / dQ^2$. We include the uncertainty coming from this as a systematic in our analysis and show the effect in Fig.~\ref{fig3}. It is seen in Fig.~\ref{3a} that, $(\overline{\nu})\nu-N$ differential cross sections determined from $G^Z_A(Q^2)$ form factor when $G^Z_A(0)$ is calculated using the lattice QCD estimate of $G^s_A(0)$, is much more precise in the low-$Q^2$ region compared to the differential cross sections shown in Fig.~\ref{3b} which are calculated from the $G^Z_A(Q^2)$ without constraining  $G^Z_A(0)$ using the lattice QCD estimate of $G^s_A(0)$. These differential cross sections at low-$Q^2$  are needed to be determined with precision. Therefore, our extraction of $(\overline{\nu})\nu-N$ differential cross sections in this low-$Q^2$ region can serve as a useful tool to numerically estimate the Pauli blocking and shadowing effects.

To demonstrate the importance of a correct determination of $G^Z_A(Q^2)$, we show in Fig.~\ref{figcontribution} that the term $\frac{G_F^2}{2\pi}\frac{Q^2}{E_\nu^2}\frac{1}{64\tau}(G^Z_A)^2W^2$ has the largest contribution to $d\sigma/dQ^2$ among individual terms in Eq.~(\ref{dsigdq}).

\begin{figure}
\begin{center}
\includegraphics[height=6.8cm,width=9.5cm]{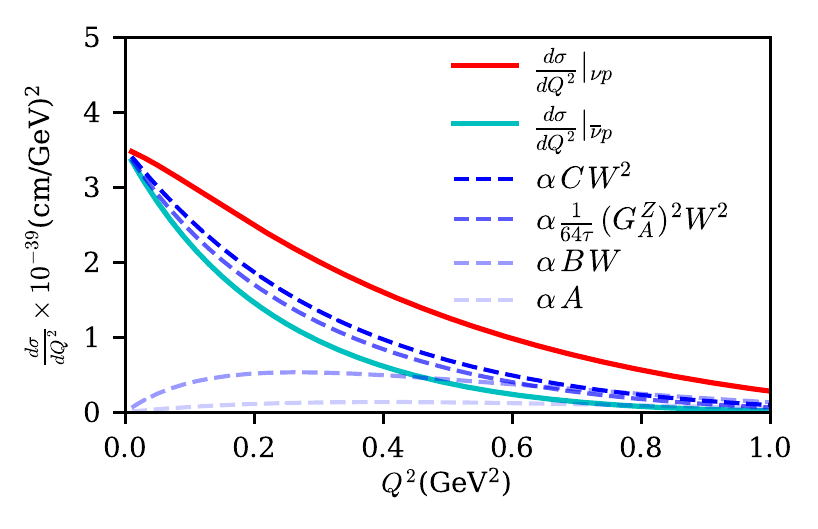}
\end{center}
\setlength\abovecaptionskip{-12pt}
\setlength\belowcaptionskip{-12pt}
\caption{Contributions from $A(Q^2)$, $B(Q^2)$, $C(Q^2)$ defined in Eq.~(\ref{defx}) and $G^Z_A(Q^2)$ to $\nu p$ and $\overline{\nu}p$ differential cross sections for neutrino beam energy $E_\nu=1$ GeV are shown as the dashed lines. The symbol $\alpha=\frac{G_F^2}{2\pi}\frac{Q^2}{E_\nu^2}$ is used in the figure for shorthand notation. The  total differential cross sections are shown with solid lines.}
\la{figcontribution}
\end{figure}
 
 \section{Prediction of BNL E734 Experimental Data}
To further test the robustness of our extraction of $G^Z_A(Q^2)$ and determination of $d\sigma/dQ^2$ for $(\overline{\nu})\nu-N$ scattering, we now describe an independent data, namely BNL E734 experiment data~\cite{Horstkotte:1981ne,Ahrens:1986xe}.  The experimental data analysis and systematics related to the BNL experiment are different to those of MiniBooNE experiments. For example, as mentioned in Ref.~\cite{AguilarArevalo:2010cx}, the MiniBooNE neutral current elastic differential cross section should be less sensitive to the final state interaction effects compared to those measured in BNL E734 experiment~\cite{Horstkotte:1981ne,Ahrens:1986xe} with tracking detectors. The BNL E734 experiment produced neutrino beam using a copper target and estimated the average neutrino and antineutrino beam energies $E_{\overline{\nu}}=1.3$ GeV and $E_\nu=1.2$ GeV respectively~\cite{Horstkotte:1981ne,Ahrens:1986xe}. Moreover, one expects that there are other sources of systematic differences between the MiniBooNE and BNL E734 published data because of various differences in the event generators, method of analyses, and so on. For this purpose, instead of using the BNL data in our analysis above, we shall predict $d\sigma_{\nu(\overline{\nu})N\to \nu(\overline{\nu})N}/dQ^2$ and compare them with those obtained from BNL E734 experiment~\cite{Horstkotte:1981ne,Ahrens:1986xe} for a given $E_\nu$. We again assume an uncertainty of $\delta E_\nu=0.1\, {\rm GeV}$  in the neutrino and antineutrino beam energy in this analysis. As shown in Fig.~\ref{fig4}, our prediction turns out to agree with the BNL E734 (anti)neutrino-nucleon scattering differential cross sections in the entire available $Q^2$-region, demonstrating the validity and predictive power of our determination of $G^Z_A(Q^2)$ and the $(\overline{\nu})\nu-N$ scattering differential  cross section using the MiniBooNE data and lattice QCD determinations of $F^s_{1,2}(Q^2)$. 
\begin{figure}
\begin{center}
\includegraphics[height=6.8cm,width=9.5cm]{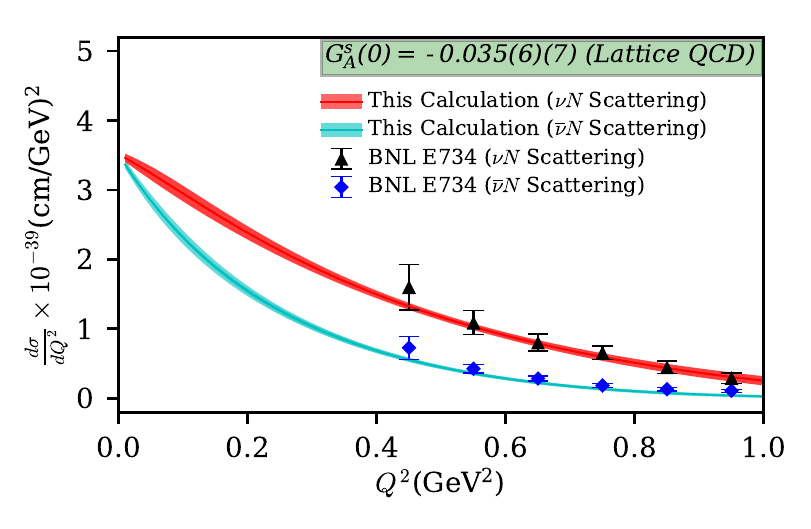}
\end{center}
\setlength\abovecaptionskip{-12pt}
\setlength\belowcaptionskip{-12pt}
\caption{Prediction of BNL E734 experiment $(\overline{\nu})\nu-N$ differential cross sections. The very mild effect coming from the neutrino beam energy $\delta E_\nu=\pm 0.1$ GeV is added as a systematic in the $d\sigma / dQ^2$ extraction and shown in the outer light red and light cyan bands. $G^Z_A(0)$ in Eq.~\eqref{gz0lat} as determined from the lattice QCD calculation of $G^s_A(0)$ and the corresponding $G^Z_A(Q^2)$ determined using the $z$-expansion fit, are used here for the prediction of $(\overline{\nu})\nu-N$ differential cross sections.}
\la{fig4}
\end{figure}

\section{Conclusion}
Using lattice calculation of strange quark axial charge, our analysis provides the most precise and model-independent determination of the neutral current weak axial charge {\mbox{$G^Z_A(0)=-0.654(3)(5)$}} which is free from nuclear effects and other systematics associated with the neutrino scattering experiments. Using this value of $G^Z_A(0)$ and the lattice QCD calculation of strange quark electromagnetic form factors in a phenomenological analysis, we also extract the most precise value of $G^Z_A(Q^2)$. Since $G^Z_A(Q^2)$ is the dominant form factor, it plays the key role in the neutral current (anti)neutrino-nucleon scattering process. The precise value of $G^Z_A(Q^2)$ and therefore the differential cross sections in this analysis constrained by the lattice QCD value of strange quark axial charge  can be used as a tool to numerically isolate nuclear effects in the low momentum transfer region of (anti)neutrino-nucleus scattering.

 Another important result of this analysis is that, although small compared to the nucleon total electromagnetic form factor, the strange quark electromagnetic form factor cannot be ignored to obtain consistent results for $G^Z_A(Q^2)$. This precise value of $G^Z_A(Q^2)$ can also help isolating higher order radiative corrections entering in the effective $G^{Z,\text{eff}}_A(Q^2)$ in parity-violating $\vec{e}-p$ scattering which are not theoretically well constrained. Finally, the robustness of the determination of the neutral current weak axial form factor is shown through the predictive power to describe $(\overline{\nu})\nu-N$ scattering differential cross sections from independent experiments. Therefore, within the limitation that we rely on the MiniBooNE published data to obtain $G^Z_A(Q^2>0)$ without additional nuclear effects incorporated in our analysis, this reliable determination of $(\overline{\nu})\nu-N$ scattering can have a significant impact in disentangling the nuclear effects in data analysis of the upcoming neutrino-nucleus scattering experiments.

\begin{acknowledgments}
  \textit{Acknowledgments:} RSS thanks Richard Hill who provided the data of the fit to nucleon electromagnetic form factors before the work was published in Ref.~\cite{Ye:2017gyb}. RSS also thanks Hans G\"unter Dosch, Guy~F.~de~T\'eramond, Joe Karpie, Luka Leskovec, Tianbo Liu,  and Aaron S. Meyer. The authors thank David Armstrong, Rajan Gupta,  Michael Kordosky, Andreas Kronfeld, Wally Van Orden, Rocco Schiavilla for useful discussions. We thank the RBC and UKQCD Collaborations for providing their DWF
gauge configurations. 

This work is supported by the U.S. Department of Energy, Office of Science, Office of Nuclear Physics under contract DE-AC05-06OR23177. This work is also supported in part by the U.S. DOE
Grant No. DE-SC0013065.
This research used resources of the Oak Ridge
Leadership Computing Facility at the Oak Ridge National Laboratory,
which is supported by the Office of Science of the U.S. Department
of Energy under Contract No. DE-AC05-00OR22725. This work used Stampede
time under the Extreme Science and Engineering Discovery Environment
(XSEDE), which is supported by National Science Foundation Grant No.
ACI-1053575. We also thank the National Energy Research Scientific
Computing Center (NERSC) for providing HPC resources that have contributed
to the research results reported within this paper. We acknowledge
the facilities of the USQCD Collaboration used for this research in
part, which are funded by the Office of Science of the U.S. Department
of Energy.

\end{acknowledgments}

\providecommand{\href}[2]{#2}
\begingroup\raggedright

\endgroup

\end{document}